\documentclass[aps,prl,twocolumn,superscriptaddress]{revtex4-1}
\usepackage{graphicx}
\usepackage{amsmath}
\usepackage{mathtools}
\usepackage{siunitx}
\usepackage{tikz}
\usepackage{amsfonts}
\usepackage{CJKutf8}
\usepackage[normalem]{ulem}
\usetikzlibrary{shapes,arrows,chains,automata, positioning}
\begin{document}


\title{Self-reinforcing directionality generates truncated L\'evy walks without the power-law assumption}



\author{Daniel Han}
\affiliation{Department of Mathematics, University of Manchester, M13 9PL, UK}

\author{Marco A. A. da Silva}
\affiliation{Faculdade de Ci\^encias Farmac\^euticas de Ribeir\~ao Preto, Universidade de S\~ao Paulo (FCFRP-USP), Ribeir\~ao Preto, Brazil}

\author{Nickolay Korabel}
\affiliation{Department of Mathematics, University of Manchester, M13 9PL, UK}

\author{Sergei Fedotov}
\affiliation{Department of Mathematics, University of Manchester, M13 9PL, UK}
\email[]{sergei.fedotov@manchester.ac.uk}

\date{\today}

\begin{abstract}
	We introduce a persistent random walk model with finite velocity and self-reinforcing directionality, which explains how exponentially distributed runs self-organize into truncated L\'evy walks observed in active intracellular transport by Chen et. al. [\textit{Nat. mat.}, 2015]. We derive the non-homogeneous in space and time, hyperbolic PDE for the probability density function (PDF) of particle position. This PDF exhibits a bimodal density (aggregation phenomena) in the superdiffusive regime, which is not observed in classical linear hyperbolic and L\'evy walk models. We find the exact solutions for the first and second moments and criteria for the transition to superdiffusion.
\end{abstract}


\maketitle
\textit{Introduction.} 
Transport in biology is crucial on multiple scales to maintain life and deviations from normal movement are hallmarks of disease and ageing \cite{murray2007mathematical,*murray2001mathematical,okubo2013diffusion}. From organisms to sub-cellular molecules, their motion is usually described by persistent random walks with finite velocities (run and tumble models) \cite{othmer1988models,hillen2002hyperbolic,tailleur2008statistical,thompson2011lattice,mendez2010reaction}. The preference for these velocity jump models over space jump random walks arise due to the physical constraints of organisms not instantaneously jumping in space and an inertial resistance to changes in direction. In recent years, L\'evy walks \cite{zaburdaev2015levy} attracted attention in modelling movement patterns of living things \cite{reynolds2018current}, from sub-cellular \cite{chen2015memoryless,song2018neuronal,fedotov2018memory,*korabel2018non} to organism \cite{raichlen2014evidence,ariel2015swarming,huda2018levy} scales. Until now, L\'evy walks have been mostly described by coupled continuous time random walks (CTRW) \cite{zaburdaev2015levy}, various fractional PDEs \cite{compte1997generalized,sokolov2003towards,meerschaert2002governing,baeumer2005space,becker2004limit,uchaikin2011fractional,magdziarz2015limit} and integro-differential equations \cite{fedotov2016single,taylor2016fractional}. These approaches require power-law distributed running times with divergent first and second moments as an \textit{ab initio} assumption. However, in many cases this assumption is difficult to justify, leading to ongoing discussions about the origin of power-law distributed runs, the Levy walk observed in nature \cite{jansen2012comment,*wosniack2017evolutionary,reynolds2018current} and L\'evy foraging hypothesis \cite{viswanathan1999optimizing,tejedor2012optimizing}.

Experiments exhibiting L\'evy-like motion cannot be modeled with pure power-law jump or running time distributions due to finite limits in physical systems \cite{reynolds2012truncated}. So theoretically, power-law distributions are often truncated \cite{mantegna1994stochastic} or exponentially tempered \cite{cartea2007fluid,*sabzikar2015tempered}. Specifically for active cargo transport in cells, it was discovered that the motion is composed of multiple short runs that self-organize into longer, truncated power-law distributed, uni-directional flights (truncated L\'evy walks) \cite{chen2015memoryless}.  Explanations for this phenomenon have been attempted in terms of \textit{self-reinforcing directionality} generated by co-operative motor protein transport  \cite{chen2015memoryless}. Yet, the question remains, can a persistent random walk model generate superdiffusion without power-law distributed runs through the self-organization of exponentially distributed runs?

In this Letter, we propose a new, spatially and temporally inhomogeneous persistent random walk model that generates exponentially truncated L\'evy walks through self-reinforced directionality. This exponential truncation is not directly introduced but rather arises naturally from the same microscopic mechanism that generates the power-law distribution of flights. Our model differs from the traditional L\'evy walk model because superdiffusion is generated by self-reinforcing directionality rather than the assumption of power-law flight distribution with infinite second moment. Simulated densities of uni-directional flight lengths from this model show excellent agreement with truncated power-law distributions observed in experiments \cite{chen2015memoryless}. It is worth noting that there are a few examples where the power-law assumption has not been used as a starting point: superdiffusion of ultracold atoms \cite{kessler2012theory,*barkai2014area} and a random walk driven by an ergodic Markov process with switching \cite{fedotov2007superdiffusion}.

\textit{Self-reinforcing directionality.}  Consider a particle moving to the right and left in 1D with constant speed $\nu$ and exponentially distributed running time with rate $\lambda$. In the standard alternating case, the particle would change directions with probability $1$. To consider the instance when there is a probability that the particle continues in the same direction as the previous run, we introduce the transition probability matrix \cite{coxmillertextbook}:

\begin{equation}
\pmb{Q} = 
\begin{bmatrix}
q_{+} & 1-q_{+}\\
1-q_{-} & q_{-}\\
\end{bmatrix},
\label{transitionmatrix}
\end{equation}
where $q_{+}$ is the conditional transition probability that the particle will continue in the positive direction given it was moving in the positive direction before. Similarly, $q_{-}$ corresponds to the particle moving in the negative direction. The standard alternating case corresponds to $q_+ = q_- = 0$.

\begin{figure}[h!]
	\centering
	\includegraphics[width=0.8\linewidth]{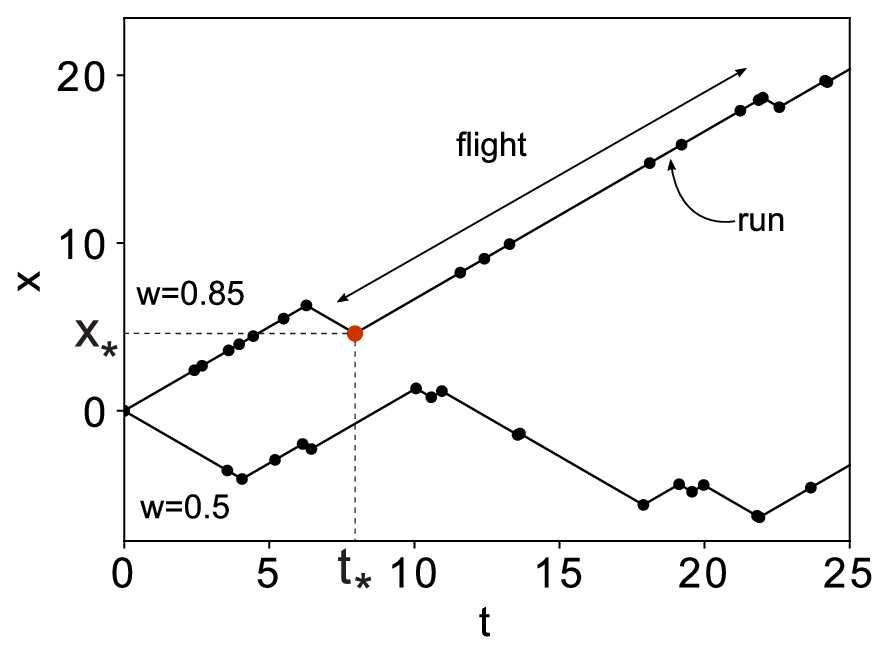}
	\caption{Two realizations of the random walk with self-reinforcing directionality using the transition matrix \eqref{transitionmatrix} with $q_{\pm}(x,t)$ from \eqref{qplus}. 
		The continuous trajectories (black solid lines) are labeled according to the persistence probability $w=0.85$ and $w=0.5$. The black dots show the beginning and end of individual exponentially distributed runs. A flight and run are annotated. The parameters are $\nu = 1$ and $\lambda = 1$. The trajectory with $w=0.85$ has an initial velocity $\nu$ and $w=0.5$, $-\nu$. The enlarged red dot shows the beginning of a uni-directional flight at $x_*$ and $t_*$.}
	\label{fig:singletrajannotated}
\end{figure}

In order to model \textit{self-reinforcing directionality} using the matrix $\pmb{Q}$, we introduce relative times, $t^+/t$ and $t^-/t$, that the particle moves in the positive and negative direction during time $t$. The key point is to define the probabilities in \eqref{transitionmatrix} as
\begin{equation}
q_{\pm} = w \frac{t^{\pm}}{t} + (1-w)\frac{t^{\mp}}{t}.
\label{qplus}
\end{equation}
Here we introduce the persistence probability, $w$, which parameterizes the extent that changes of direction are affected by relative times. If $w=1/2$ then the transition probabilities in both directions are the same: $q_+ = q_- = 1/2$. If $w>1/2$ then the matrix $\pmb{Q}$ has \textit{repetition compulsion} properties: the longer a particle spends moving in the one direction, the greater the probability to maintain directionality. Figure \ref{fig:singletrajannotated} illustrates self-organization of individual runs into long uni-directional flights. 

Let us show that the conditional transition probabilities \eqref{qplus} can be expressed in terms of particle position $x= x_0+\nu\left[t^+-t^-\right]$. Since $t=t^++t^-$, the relative times can be written as
\begin{equation}
	\frac{t^{\pm}}{t} = \frac{1}{2}\left(	1\pm\frac{x-x_0}{\nu t}	\right).
	\label{tplus}
\end{equation}
Then substituting \eqref{tplus} into \eqref{qplus}, we find that the probabilities $q_{+}$ and $q_{-}$ depend on $x$ and $t$,
\begin{equation}
q_{\pm}(x,t) = \frac{1}{2}\left[1\pm\alpha\frac{x-x_0}{\nu t}\right]\text{ with } \alpha = 2w-1.
\label{qplusxt}
\end{equation}
Since our random walker is moving with finite speed $\nu$, the relation $(x-x_0)/\nu t \leq 1$ must hold. Once again, the formula \eqref{qplusxt} shows self-reinforcing directionality since if $\alpha>0$ ($w>1/2$) then the transition probability $q_{+}(x,t)$ is an increasing function of particle position $x$ and the opposite for $q_-(x,t)$.

The equations for probability density functions (PDFs) of particles moving right ($+$) and left ($-$), $p_{\pm}(x,t)$, can be written in terms of transition matrix $\pmb{Q}$ as
\begin{equation}
\begin{split}
\begin{bmatrix}
p_{+}(x,t+\Delta t)\\
p_{-}(x,t+\Delta t)
\end{bmatrix}
&=(1-\lambda\Delta t)
\begin{bmatrix}
p_{+}(x-\nu\Delta t,t)\\
p_{-}(x+\nu\Delta t,t)
\end{bmatrix} \\
& +\pmb{Q}^{T}
\begin{bmatrix}
p_{+}(x,t)\\
p_{-}(x,t)
\end{bmatrix} \lambda\Delta t.
\end{split}
\label{diffeq}
\end{equation}
Rearranging and taking the limit $\Delta t\rightarrow0$, we can write the equations for $p_{\pm}(x,t)$ as
\begin{equation}
\begin{split}
\frac{\partial p_{\pm}}{\partial t}\pm\nu\frac{\partial p_{\pm}}{\partial x} &= -\lambda(1-q_{\pm}(x,t)) p_{\pm}+ \lambda(1-q_{\mp}(x,t) )p_{\mp}.
\end{split}
\label{master2}
\end{equation}
Note these equations \eqref{master2} can be rewritten in terms of space and time dependent switching rates
\begin{equation}
\lambda_{\pm}(x,t) = \lambda(1-q_{\pm}(x,t)).
\label{effectiverate}
\end{equation}These switching rates will be used to show how the exponentially truncated power-law distribution of flights arise from exponentially distributed runs. Using standard methods \cite{mendez2010reaction,schnitzer1993theory,thompson2011lattice} and equation \eqref{qplusxt}, we can obtain the partial differential equations (PDEs) for total density $p(x,t) = p_{+}+p_{-}$ and flux $J(x,t) = \nu p_+-\nu p_-$: $\partial p/\partial t = -\partial J/\partial x$ and $\partial J/\partial t = -\nu^2 \partial p/\partial x -\lambda (J - \alpha x p/t )$ (See Supp. Mat.). The initial conditions are
\begin{equation}
\begin{split}
	p(x,0) = \delta(x-x_0), \text{ } J(x,0) = \nu (2u-1) \delta(x-x_0),
	\label{initialconditions}
\end{split}
\end{equation}
where $u \in [0,1]$ is the probability that the intial velocity is $\nu$. Finally, we can find the hyperbolic PDE with a non-homogeneous in space and time advection term
\begin{equation}
\begin{split}
\frac{\partial^2 p}{\partial t^2} + \lambda\frac{\partial p}{\partial t} = \nu^2 \frac{\partial^2 p}{\partial x^2} - \frac{ \lambda\alpha}{t} \frac{\partial ((x-x_0) p)}{\partial x}, \text{ } t>0.
\end{split}
\label{master}
\end{equation}

The advection term of Eq. \eqref{master} is unconventional because it depends on the initial position $x_0$. Furthermore, if the initial conditions are symmetric, $u=1/2$ (see \eqref{initialconditions}), then the average drift is zero. Clearly, \eqref{master} is a modification of the classical telegraph or Cattaneo equation \cite{kac1974stochastic,schnitzer1993theory,mendez2010reaction}, with a time and space dependent advection term. In what follows, we will show that this additional term generates superdiffusion. In fact, a generalized Cattaneo equation generating superdiffusion has been formulated using the Riemann-Liouville fractional derivative \cite{compte1997generalized}. The advantage of Eq. \eqref{master} over fractional PDEs is that it is far simpler and does not require integral operators in time. To the authors' knowledge, the hyperbolic PDE \eqref{master} is the first formulation of truncated L\'evy walks without integral operators \cite{fedotov2016single}. 
In the diffusive limit, when $\lambda\rightarrow\infty$ and $\nu\rightarrow\infty$ such that $\nu^2/\lambda$ is a constant, \eqref{master} becomes the governing advection-diffusion equation for the continuous approximation of the elephant random walk \cite{schutz2004elephants,*paraan2006exact,*da2013non,*da2020non}. Note that the system of equations \eqref{master2} with transition rates \eqref{qplusxt} can also be mapped to the hyperbolic model for chemotaxis \cite{hillen2000hyperbolic,*stevens1997aggregation} with an unorthodox external stimulus $S(x,t)$ (see Supp. Mat.)

\textit{Moment analysis.} Now, we show that the variance for the underlying random process, $x(t)$, exhibits superdiffusive behavior: $\text{Var}\{x(t)\} \propto t^{2\alpha}$ with $1/2<\alpha<1$. The moments of random walk position, $\mu_n(t) = \int_{-\infty}^{\infty}x^np(x,t)dx$, can be found from \eqref{master} as
\begin{equation}
\begin{split}
\frac{d^2 \mu_n}{dt^2} - \nu^2 n (n-1) \mu_{n-2} + \lambda \frac{d\mu_n}{dt} - \frac{\lambda\alpha n}{t}\mu_n = 0.
\end{split}
\label{momentmaster}
\end{equation}
Taking the Laplace transform and solving the resulting differential equations for the first and second moments using the initial conditions, $\mu_1(0) = \mu_2(0) = 0$, $ d\mu_1(0)/dt =\nu(2u-1)$, $ d\mu_2(0)/dt= 0$ we can obtain  $\hat{\mu}_1(s) = \nu(2u-1) s^{-1-\alpha}(s+\lambda)^{\alpha-1}$ and $\hat{\mu}_2(s) = \frac{2\nu^2}{\lambda(2\alpha-1)}\left[ s^{-1-2\alpha}(s+\lambda)^{2\alpha-1}-s^{-2} \right]$ (for details see Supp. Mat.).
The inverse Laplace transform gives $\mu_1(t) = \nu(2u-1)t \prescript{}{1}{F}_1(1-\alpha,2,-\lambda t)$ and $\mu_2(t) = 2\nu^2 t \left[\prescript{}{1}{F}_1(1-2\alpha,2,-\lambda t) -1 \right]/(\lambda(2\alpha -1))$, where $\prescript{}{1}{F}_1(a,b,z)$ is the Kummer confluent hypergeometric function. In the long time limit ($s\rightarrow0$), we obtain
\begin{equation}
	\mu_1(t) \simeq \frac{\nu \lambda^{\alpha-1}(2u-1)}{\Gamma(\alpha+1)}t^{\alpha}
	\label{firstmomentsolnasymp}
\end{equation}
and 
\begin{equation}
	\mu_2(t) \simeq 
	\begin{cases}
	\frac{2\nu^2}{\lambda(1-2\alpha)} t, & -1<\alpha<1/2 \\
	\frac{2\nu^2\lambda^{2\alpha-2}}{(2\alpha-1)\Gamma(2\alpha+1)}t^{2\alpha}, & 1/2<\alpha <1
	\end{cases}
	\label{secondmomentsolnasymp}
\end{equation}
Clearly, the random walk exhibits superdiffusive behavior for $1/2<\alpha<1$.
The variance, $\text{Var}\left\{x(t)\right\}= \mu_2(t) - \mu_1(t)^2$, is
\begin{equation}
\text{Var}\left\{x(t)\right\} \simeq 
\begin{cases}
\frac{2\nu^2}{\lambda(1-2\alpha)}t, & -1<\alpha<1/2 \\
A\nu^2\lambda^{2\alpha-2} t^{2\alpha}, & 1/2<\alpha <1
\end{cases}
\label{varianceasymp}
\end{equation}
where $A = \frac{2}{(2\alpha -1)\Gamma(2\alpha+1)} - \frac{(2u-1)^2}{\Gamma^2(\alpha+1)}$. 
Results of Monte Carlo simulations are in perfect agreement with the analytical solution of the second moment (see Supp. Mat.). 

\textit{Truncated L\'evy walk.} In what follows, we show analytically how self-reinforcing directionality generates a truncated L\'evy walk with exponentially tempered power-law distributed flights. Let us find the conditional distribution of flights given the particle changes direction at the position $x_*$ and time $t_*$ (illustrated in Fig. \ref{fig:singletrajannotated} where the particle changes velocity from $-\nu$ to $\nu$). First, we calculate the conditional survival function, $\Psi_{\pm}(\tau| x_*,t_*)$, for flights moving in the positive ($+$) and negative ($-$) direction. This function gives us the probability that the random duration of a flight, $T$, is greater than $\tau$. The rates, $\lambda_{\pm}$ \eqref{effectiverate} define the conditional switching rate along the particle trajectory starting at position $x_*$ and time $t_*$ as $\lambda_{\pm}(\tau|x_*,t_*) = \lambda(1-q_{\pm}(x_*\pm\nu\tau,t_*+\tau))$. If we rearrange using $\alpha = 2w-1$, then 
\begin{equation}
\lambda_{\pm}(\tau| x_*,t_*) = \lambda (1-w) + \frac{\gamma_{\pm}}{t_* + \tau},
\label{condswitchrate2}
\end{equation}
where $\gamma_{\pm} = (w-\frac{1}{2})\lambda \left( t_* \mp\frac{x_*}{\nu} \right)$. Using this, we can find the conditional survival function by $\Psi_{\pm}(\tau | x_*,t_*) = \exp\left(-\int_{0}^{\tau}\lambda_{\pm}(s| x_*,t_*)ds\right)$. Then, we can see that the constant term in \eqref{condswitchrate2} produces an exponential tempering factor and the term inversely proportional to the running time, $\tau$, generates a power-law with exponent $\gamma_{\pm}$. Explicitly, we can write
\begin{equation}
\Psi_{\pm}(\tau| x_*,t_*) = e^{-(1-w)\lambda\tau}\left(\frac{t_*}{t_*+\tau}\right)^{\gamma_{\pm}}.
\label{conditionalsurv_explicit}
\end{equation}
The conditional distribution of flight lengths $l=\nu\tau$ is given by: $F_{\pm}(l|x_*,t_*) = 1- \Psi_{\pm}(l/\nu| x_*,t_*) $. Each uni-directional flight is drawn from a truncated power-law distribution dependent on the position $x_*$ and time $t_*$, where the particle changes direction. Clearly, this demonstrates the difference between our new model and the traditional L\'evy walk model where the power-law distribution of flights is an \textit{a priori} requirement and not dependent on $x_*$ and $t_*$. Furthermore, our new model is able to generate superdiffusion \eqref{secondmomentsolnasymp} without an infinite second moment of flight times \eqref{conditionalsurv_explicit} through self-reinforcing directionality, unlike traditional L\'evy walk models.

In addition, exponential truncation (tempering) is  usually introduced by simply multiplying the power-law jump or waiting time densities by an exponential factor with an additional parameter, leading to tempered fractional calculus \cite{cartea2007fluid,*sabzikar2015tempered}. The advantage of our model is that both exponential tempering and power-law flight distribution are generated through a single microscopic mechanism involving \textit{self-reinforced directionality} and the subsequent analysis of \eqref{master} is more convenient than tempered fractional calculus. An interesting feature of our model is that despite the exponential truncation of flights, the variance \eqref{varianceasymp} still exhibits superdiffusive behavior for $1/2<\alpha<1$ ($3/4<w<1$). Numerical simulations confirm exponentially truncated power-law conditional survival function \eqref{conditionalsurv_explicit} for flight times (see Supp. Mat.).

\begin{figure}[h!]
	\centering
	\includegraphics[width=0.8\linewidth]{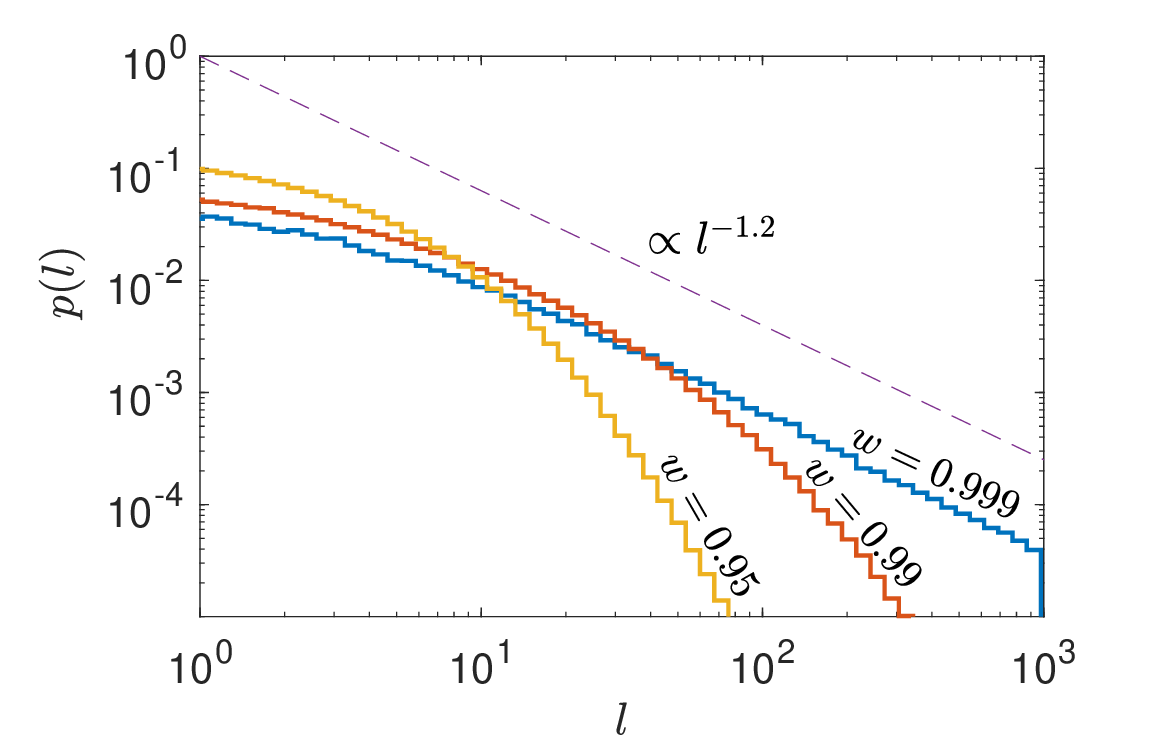}
	\caption{The PDFs $p(l)$ (solid lines) of unidirectional flight lengths $l$ from the simulations of $N= 10^5$ particles with varying $w$, $t=10^3$, $v=1$ and $\lambda =1$.}
	\label{flights_pdf}
\end{figure}
	
Finally, we can obtain numerically the PDF, $p(l)$, of flight length, $l$, without the condition that flights begin at $x_*$ and $t_*$. In the case of strong directionality as $w\rightarrow1$, $p(l)$ should approach a pure power-law since the truncation length $\nu/\lambda(1-w) \rightarrow \infty$. Figure \ref{flights_pdf} confirms this and shows that, when $w = 0.999$, a power-law density is recovered for more than two orders of magnitude. Next, we provide evidence that the truncated power-law PDF shows excellent correspondence to published data on uni-directional endosome flights.

\begin{figure}
	\centering
	\includegraphics[width=0.8\linewidth]{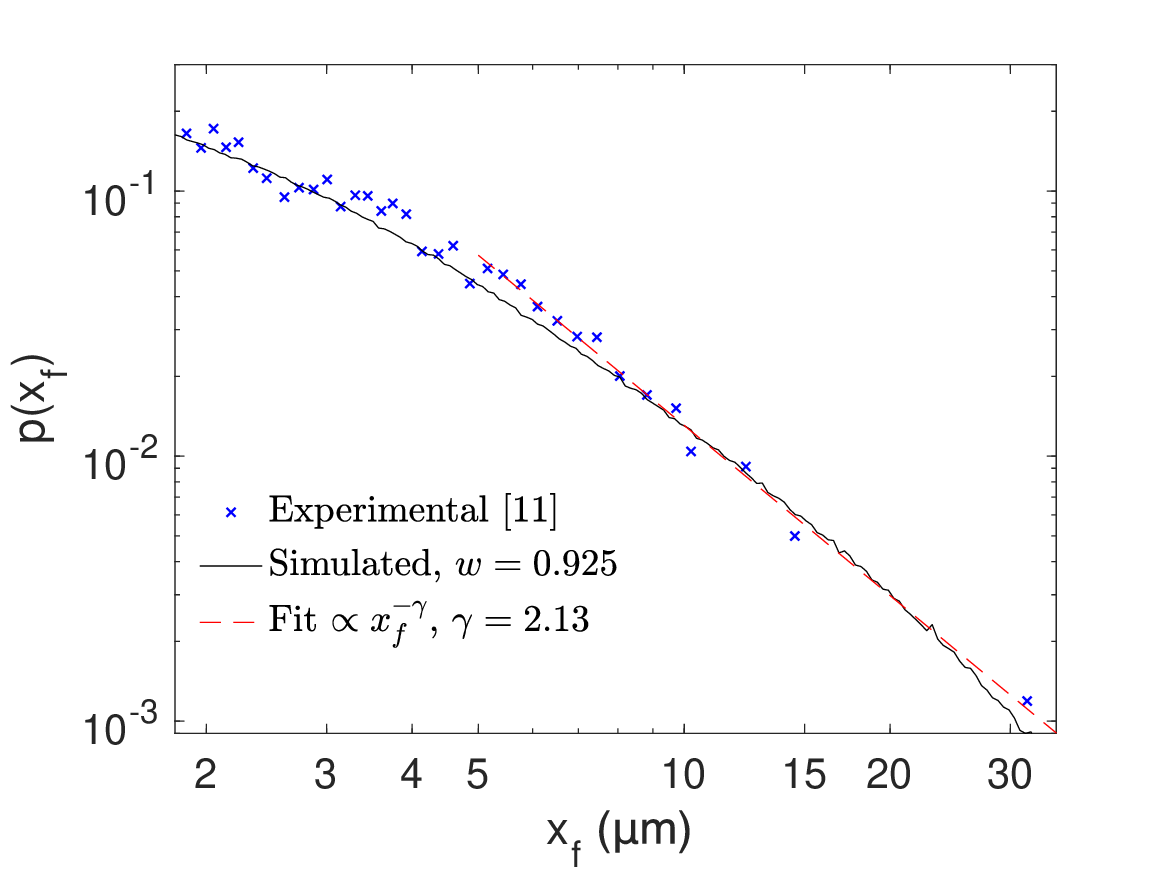}
	\caption{The PDFs, $p(x_{f})$, of uni-directional flight lengths, $x_{f}$ from experimental measurements (crosses) taken from \cite{chen2015memoryless} and simulation of underlying random walk with $w=0.925$ (solid black). Each simulation had $N=10^5$ particles running for $t=10^2$\SI{}{\second} with $\nu=$ \SI{1.2}{\micro\meter\per\second} and $\lambda=$\SI{1}{\per\second}. The tail of the PDF was fitted with a power-law (red dashed), $p(x_f)\propto x_f^{-\gamma}$ with $\gamma = 2.13$. 
	}
	\label{fig:flightdisp_histo}
\end{figure}

\textit{Experimental data and numerical simulations.} Truncated L\'evy walk behavior is observed experimentally in intracellular transport \cite{chen2015memoryless}, which arises from the self-organization of exponentially distributed runs, $x_i$, into uni-directional flights, $x_f$ \cite{chen2015memoryless}. Here, we switch from $l$ to $x_f$ to avoid confusion between theoretical and experimental flights. Until now, there had been no governing PDE like \eqref{master} and an underlying persistent random walk model to describe this phenomenon. Experimentally, the authors \cite{chen2015memoryless} report power-law tails in the flight-length density, $p(x_f)\propto x_f^{-2}$. Figure \ref{fig:flightdisp_histo} demonstrates that numerical simulations of our model are able to generate the power-law tails for flight length density and emulate the experimental data on the whole $x_f$ scale using reasonable parameters. Furthermore, the parameters of our new model, such as persistence probability $w$ or $\alpha$, rate $\lambda$ and speed $\nu$ can be easily found by comparing the exact analytical formula of second moment with experimental mean squared displacements.

\begin{figure}
	\includegraphics[width=0.8\linewidth]{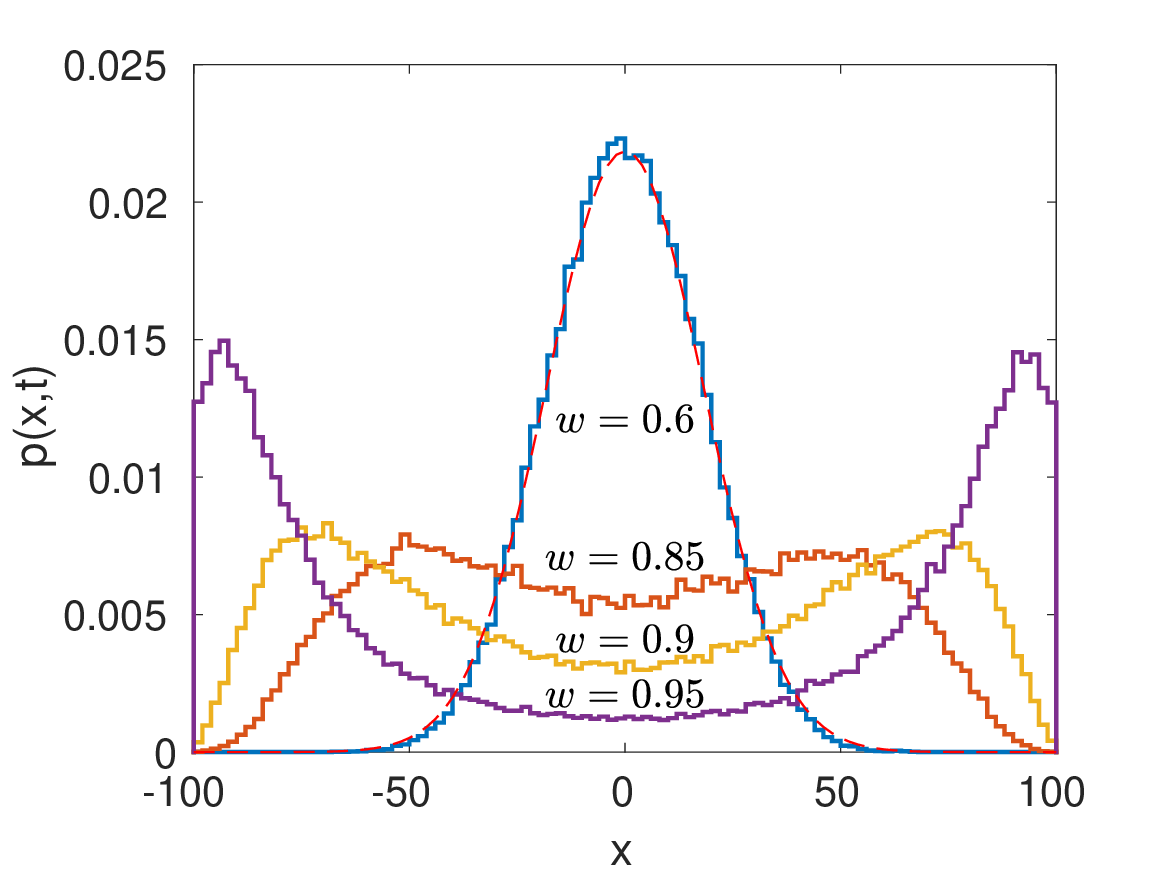}
	\caption{The PDFs (solid lines) $p(x,t)$ for particle positions $x$ at the end of simulation time $t=10^2$ for varying values of $w$. The Gaussian solution (dashed line) is shown for $w=0.6$. The parameters of simulation were $N=10^5$ particles, $u=0.5$, $\lambda=1$, $\nu=1$. The initial density was $p(x,0) = \delta(x)$.}
	\label{poshistogram}
\end{figure}

\textit{Bimodal densities and transition from diffusive to superdiffusive regime.} Surprisingly, our truncated L\'evy walk model in the long time limit leads to bimodal densities in the superdiffusive regime ($1/2<\alpha<1$). This phenomenon does not exist for classical superdiffusive L\'evy walks. Bimodal densities (Lamperti distributions) only appear in the ballistic regime for L\'evy walks with a divergent first moment for running times \cite{uchaikin2009statistical,klafter2011first,uchaikin2011fractional,zaburdaev2015levy,froemberg2015asymptotic}. In the superdiffusive case, the density for L\'evy walks is Gaussian in the central part with power-law tails \cite{klafter2011first}. This is completely different to densities for \eqref{master} (see Fig. \ref{poshistogram}). Density peaks for $p(x,t)$ in \eqref{master} occur at $|x|<\nu t$. Figure \ref{poshistogram} shows the emergence of a bimodal density for $t=10^2$ and $w>w_c$, where $w_c=3/4$. The density $p(x,t)$ exhibits two distinct long time behaviors: it is Gaussian for $\alpha<1/2$ ($w<w_c$) and bimodal for $\alpha>1/2$ ($w>w_c$). The form of the PDF also shows a non-equilibrium phase transition (see Supp. Mat.). Note that similar bimodal densities are observed in velocity random walks with interacting particles \cite{lutscher2002emerging,*fetecau2010investigation,*carrillo2014non,fedotov2017emergence}.


For $-1<\alpha<1/2$ ($0<w<3/4$), the variance \eqref{varianceasymp} corresponds to the diffusive regime with the effective diffusion coefficient $D = \nu^2/\lambda(1-2\alpha)$. For alternating velocity random walks, the conventional diffusion coefficient would be $D_0=\nu^2/\lambda$. As $\alpha\rightarrow1/2$, the effective diffusion coefficient tends to infinity and this indicates the transition from the diffusive to superdiffusive regime. For the long time limit of \eqref{master} in the diffusive regime when $\partial^2 p/\partial t^2$ becomes negligible, the solution is Gaussian: $p(x,t) = (4\pi D t)^{-1/2}\exp\left(-(x-\mu_{\alpha}t^{\alpha})^2/4Dt\right),$	where $\mu_{\alpha} =\nu(2u-1)\lambda^{\alpha-1}/\Gamma(\alpha+1)$. Figure \ref{poshistogram} shows the excellent agreement between numerical simulations and the Gaussian solution for $w=0.6$.

\textit{Self-reinforced directionality in endosome movement.} Now, the question remains: what is the underlying biological mechanism explaining self-reinforced directionality in intracellular transport? To answer this question, we suggest a simple illustration of the possible origin for self-reinforced directionality. Let us consider endosomal motility, which is governed by the adaptor complexes on its membranes, the most notable being Rab GTPases \cite{stenmark2009rab}. These adaptors facilitate attachment to dynein and kinesin motors \cite{urnavicius2018cryo} and, therefore, dictate the positioning and motility of endosomes in the cell \cite{stenmark2009rab,gindhart2009lysosome}. To simplify this vastly complex process, consider that the endosome contains a number, $n_-$, of adaptor proteins that attach to kinesin leading to transport towards the cell periphery. Similarly, $n_+$ is the number of adaptor proteins that attach to dynein and facilitate transport towards the cell nucleus. Then, when an endosome happens to attach to a microtubule, the simplest assumption about probabilities $q_+$ and $q_-$ in \eqref{transitionmatrix} would be $q_+= n_{+}/(n_++n_-)$ and $q_-= n_{-}/(n_++n_-)$. However, due to the complexity of endosomal transport we can introduce a weight, $w \in [0,1]$, such that $q_{\pm} = wn_{\pm}/(n_++n_-) + (1-w)n_{\mp}/(n_++n_-).$

From the very beginning of endocytosis until degradation, endosomes undergo a maturation process, including the association of proteins, such as Rab5 and PI(3)K \cite{huotari2011endosome}. Recent work has show that effectors of Rab5 display distinct spatial densities \cite{villasenor2016signal} suggesting that $n_-$ and $n_+$ are functions of the time spent running towards, $t^-$, or away, $t^+$, from the cell center. So, we assume that $n_{\pm}=n_{\pm}^0 + at^{\pm} $ with $a$ being some constant rate. The more an endosome moves in towards the nucleus, the more $n_+$ increases and vice versa. This reinforcement rule is similar to that of discrete reinforced random walks \cite{davis1990reinforced,stevens1997aggregation}. Neglecting $n_{\pm}^0$, this formulation is exactly what leads to the repetition compulsion property in \eqref{qplus}, since then $n_-/(n_++n_-) = t^-/t$ and $n_+/(n_++n_-) = t^+/t$.


\textit{Summary.} In this Letter, we developed a persistent random walk model with finite velocity that generates superdiffusion without the \textit{ab initio} assumption of power-law distributed run times with infinite second moment. Our model shows that truncated power-law distributed flights can be generated from exponentially distributed runs through self-reinforcing directionality. A governing hyperbolic PDE \eqref{master} for particle probability density was derived along with exact solutions for the first and second moments. The theory is able to explain the experimentally observed self-organization of exponentially distributed runs into unidirectional flights leading to exponentially truncated L\'evy walks \cite{chen2015memoryless}. We showed excellent agreement between the density of flight lengths from numerical simulations and \textit{in vivo} cargo transport experiments. In the superdiffusive regime, numerical simulations of particle densities show bimodal densities (aggregation), which is a new phenomenon not seen in the classical linear hyperbolic or L\'evy walk models. We believe that our methodology can be used to model migrating cancer cells \cite{mierke2011integrin,*mierke2013integrin,huda2018levy}, T-cell motility \cite{harris2012generalized}, human foraging \cite{raichlen2014evidence}, front propagation phenomena \cite{mendez2004front,*fedotov2015persistent}, first passage time problems \cite{redner2001guide,condamin2007first,*angelani2014first}and viruses mobility inside cells \cite{lagache2008effective,*greber2006superhighway,*dodding2011coupling}. 




\begin{acknowledgments}
	The authors acknowledge financial support from FAPESP/SPRINT Grant No. 15308-4 , EPSRC Grant No. EP/J019526/1 and the Wellcome Trust Grant No. 215189/Z/19/Z. We thank S. Granick for sharing experimental data; V. J. Allan, T. A. Waigh and P. Woodman for discussion.
\end{acknowledgments}

\bibliography{real}

\renewcommand{\theequation}{S.\arabic{equation}}
\renewcommand\thefigure{S\arabic{figure}} 
\section*{Supplementary Material}

\subsection{Derivation of the hyperbolic PDE}

From \eqref{master2}, we can add and subtract the equations to obtain
\begin{equation}
\frac{\partial (p_{+}+p_{-})}{\partial t}  +\nu\frac{\partial (p_{+}-p_{-})}{\partial x} = 0
\end{equation}
and
\begin{equation}
\begin{split}
\frac{\partial (p_{+}-p_-)}{\partial t}+\nu\frac{\partial (p_{+}+p_-)}{\partial x}&=\\ -2\lambda(1-q_{+}(x,t)) p_{+}+& 2\lambda(1-q_{-}(x,t) )p_{-}.
\end{split}
\end{equation}
Then using definitions of the total probability $p(x,t) = p_+(x,t) + p_-(x,t)$, the flux $J(x,t) = \nu \left[p_+(x,t) - p_-(x,t)\right]$ and conditional transition probabilities $q_{\pm}(x,t) = \frac{1}{2}\left[1\pm\alpha\frac{x-x_0}{\nu t}\right]$, we can write
\begin{equation}
\frac{\partial p}{\partial t} + \frac{\partial J}{\partial x} = 0
\label{consveq}
\end{equation}
and
\begin{equation}
\frac{1}{\nu}\frac{\partial J}{\partial t} + \nu \frac{\partial p}{\partial x} = -\frac{\lambda}{\nu}J + \lambda \alpha \frac{(x-x_0)}{\nu t} p.
\label{fluxeq}
\end{equation}
Then by differentiation, \eqref{consveq} and \eqref{fluxeq} become,
\begin{equation}
\frac{\partial^2 p}{\partial t^2} = -\frac{\partial^2 J}{\partial x \partial t}
\label{2consveq}
\end{equation}
and
\begin{equation}
\frac{\partial^2 J}{\partial t \partial x} = -\nu^2 \frac{\partial^2 p}{\partial x^2} - \lambda \frac{\partial J}{\partial x} +\frac{\lambda\alpha}{t}\frac{\partial  \left((x-x_0)p\right)}{\partial x}.
\label{2fluxeq}
\end{equation}
Then combining \eqref{2consveq} and \eqref{2fluxeq}, we obtain the hyperbolic PDE \eqref{master}.

\section{Simulations}
The simulations of our correlated random walk is as follows:
\begin{enumerate}
	\item Set initial conditions $x_0=0$ and $t_0=0$. For initial velocity draw a uniformly distributed random number $U\in[0,1)$, if $U<u$ then $v_0=\nu$ and otherwise $v_0 = -\nu$.
	\item Generate an exponentially distributed random time $T_0 = -1/\lambda \log(1-V)$ where $V$ is a uniformly distributed random number in $[0,1)$. 
	\item Update position and time to $x_1 = x_0 + v_0 T_0$, $t_1 = t_0 + T_0$ respectively. For updating velocity, draw a uniformly distributed random number $W\in[0,1)$, then
	\begin{enumerate}
		\item If $v_0=\nu$ and $W<q_-(x_1,t_1) = 1/2 - \alpha (x_1/2 \nu t_1)$ then $v_1 = -\nu$. 
		\item If $v_0=-\nu$ and $W<q_+(x_1,t_1) = 1/2 + \alpha (x_1/2 \nu t_1)$ then $v_1 = \nu$. 
	\end{enumerate}
	\item Repeat steps 2 and 3 until $t_n = t_0 +\sum_{i=0}^{n-1}T_i$ reaches the end of the simulation time $t_{end}$.
\end{enumerate}

\begin{figure}[h!]
	\includegraphics[width=\linewidth]{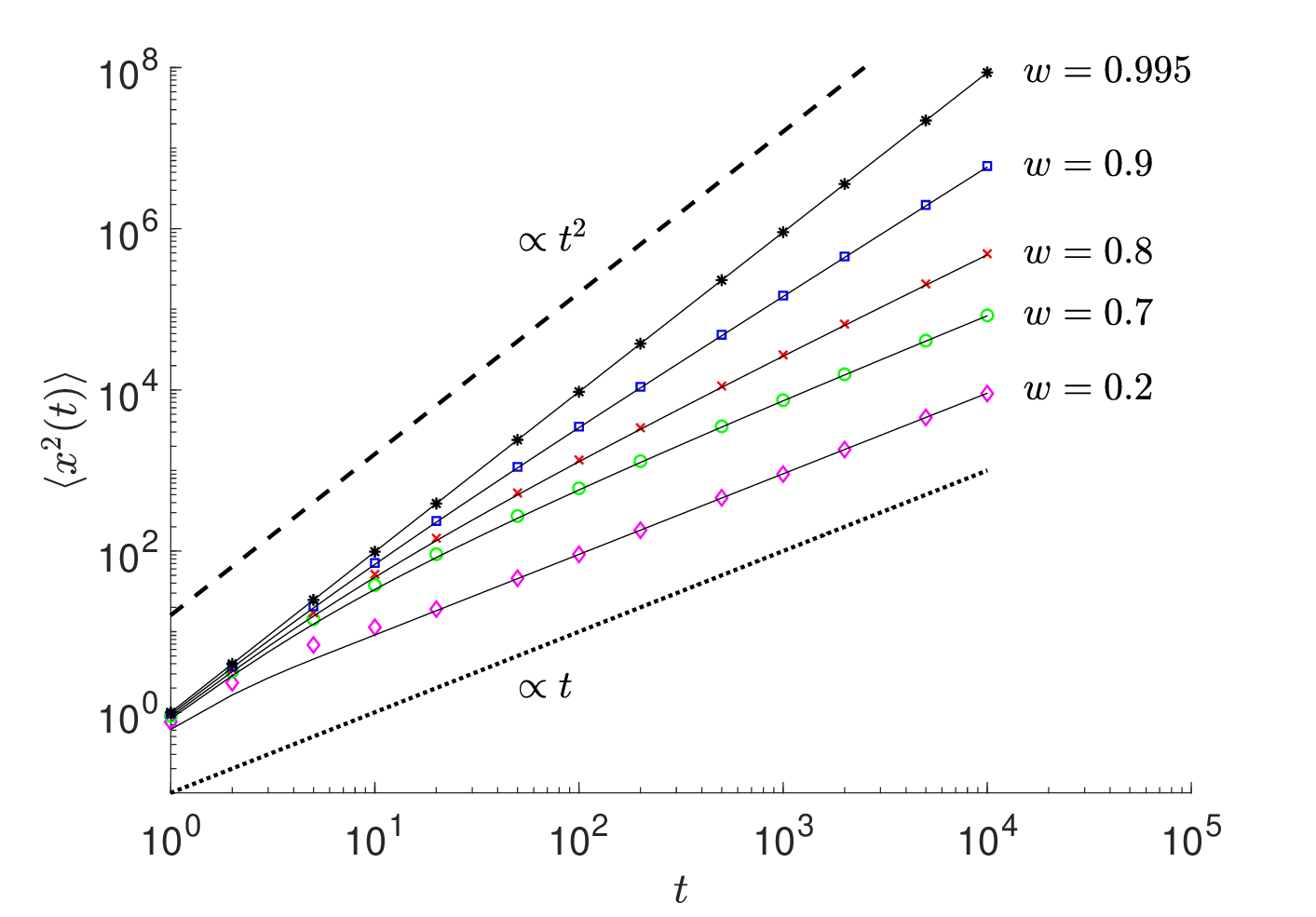}
	\caption{Mean squared displacements, $\langle x^2(t)\rangle = \mu_2(t)$, from simulated trajectories (data points) compared with the analytical solutions (solid lines). Each pair is annotated with the simulation value of $w$. Each simulation, contained $N=10^5$ particles that ran for a simulation time $t=10^4$ with parameters $u=1$,$\nu=1$ and $\lambda=1$. Diffusion (dotted line), $\langle x^2(t)\rangle \sim t$, and ballistic motion (dashed line), $\langle x^2(t)\rangle \sim t^2$, are also shown.}
	\label{secondmomentsimulations}
\end{figure}

\section{Moment calculations}
From Eq. \eqref{momentmaster}, the first moment, $\mu_1(t)$, obeys
\begin{equation}
\frac{d^2\mu_1(t)}{dt^2} + \lambda \frac{d\mu_1(t)}{dt} -  \frac{\lambda\alpha}{t}\mu_1(t) = 0
\label{firstmoment}
\end{equation}
and the second moment, $\mu_2(t)$,
\begin{equation}
\frac{d^2 \mu_2(t)}{dt^2} + \lambda \frac{d\mu_2(t)}{dt}- \frac{2\lambda\alpha}{t}\mu_2(t) = 2\nu^2.
\label{secondmoment}
\end{equation} 

We take the Laplace transform of \eqref{firstmoment} and \eqref{secondmoment}, which gives us differential equations for $\hat{\mu}_n(s) = \int_{0}^{\infty}\mu_n(t)e^{-st}dt$, for $n=1$ 
\begin{equation}
\frac{d\hat{\mu}_1(s)}{ds} = -\frac{2s + \lambda + \lambda \alpha}{s^2 + \lambda s}\hat{\mu}_1(s) 
\label{laplacefirstmoment}
\end{equation}
and $n=2$
\begin{equation}
\frac{d\hat{\mu}_2(s)}{ds} = -\frac{2s + \lambda + 2\lambda \alpha}{s^2 + \lambda s}\hat{\mu_2}(s) - \frac{2\nu^2}{s^2(s^2+\lambda s)}.
\label{laplacesecondmoment}
\end{equation}
Then, solving \eqref{laplacefirstmoment} and \eqref{laplacesecondmoment} with initial conditions we obtain the analytical solutions for the first and second moments in the Letter.

To verify our analytical results for $\mu_1(t)$ and $\mu_2(t)$, we performed simulations of random walks governed by (9). From these simulations, we can measure the ensemble average mean squared displacement $\langle x^2(t) \rangle = \mu_2(t)$. Figure \ref{secondmomentsimulations} shows excellent agreement between the analytical solutions and numerical simulations. This clearly demonstrates the emergence of superdiffusion since for $w<3/4$ ($\alpha<1/2$), $\mu_2(t) = \langle x^2(t) \rangle \propto t$, whereas $\langle x^2(t) \rangle \propto t^{2\alpha}$ for $w>3/4$ ($\alpha>1/2$).

\section{Link with Chemotaxis}
It is interesting to note that the system of equations (6) with transition rates (4) can be mapped to the hyperbolic model for chemotaxis with an unorthodox external stimulus $S(x,t)$ obeying the Hamilton-Jacobi equation for a free particle. In terms of effective turning rates, $\lambda_{\pm}$ introduced in (7), that depend on the gradient of external stimulus: $\lambda_{\pm}\left(S_x\right) = (\lambda/2) \left[1\mp\frac{1}{\nu} S_x \right]$, with $S(x,t) = \frac{\alpha(x-x_0)^2}{2t}$. Then the Hamilton-Jacobi equation for the external stimulus is $\partial S/\partial t + \frac{1}{2} \left(\partial S/\partial x\right)^2 = 0$. This provides insight into how the external stimulus, $S$, generates superdiffusion rather than the conventional ballistic motion in chemotaxis. 

\section{Truncated power law flight simulations}

\begin{figure}[h!]
	\centering
	\includegraphics[width=0.8\linewidth]{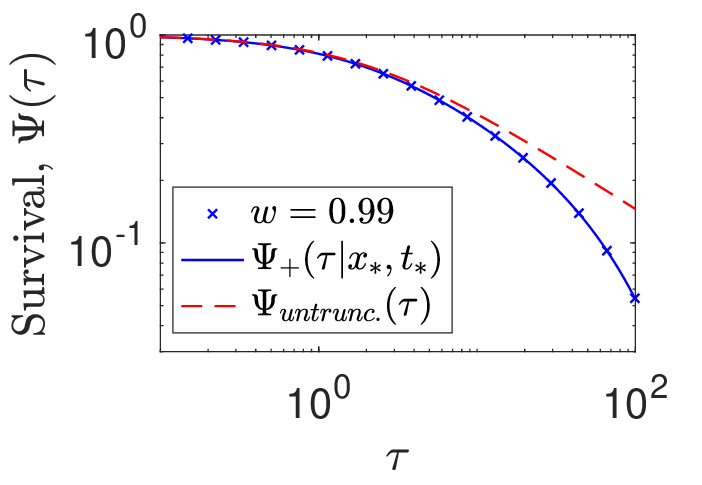}
	\caption{ The survival function $\Psi(\tau)$ of unidirectional flight running times $\tau$, estimated using the non-parametric Kaplan-Meier method, from the simulation of $N=2\times 10^6$ unidirectional flights with $w=0.99$, $v=1$, $\lambda =1$, $t_*=2$ and $x_*=1$. The conditional survival function of flights from numerical simulations (blue crosses) and the analytical truncated power-law conditional survival function $\Psi_{+}(\tau | x_*,t_*)$ from (20) (blue line) show excellent correspondence. The survival function of  $\Psi_{+}(\tau | x_*,t_*)$ without the exponential truncation, as defined in \eqref{untrunc}, is shown for comparisons (red dashed line). }
	\label{flights_surv}
\end{figure}

Fig. \ref{flights_surv} shows the evidence of exponentially truncated power-law conditional survival functions and the comparison between numerical simulations and the analytical formula (20). In addition, we compare the truncated power-law survival function, $\Psi_{+}(\tau | x_*,t_*)$ from (20), to its untruncated power-law form:
\begin{equation} 
\Psi_{untrunc.}(\tau) = \left(\frac{t_*}{t_* + \tau}\right)^{\gamma_{+}}.
\label{untrunc}
\end{equation}

\section{Decaying fronts at the propagation limit}

The bimodal distribution of $p(x,t)$ in Fig. 4 with peaks close to the maximum position $\pm \nu t$ is reminiscent of the delta function horns at $x=\pm \nu t$ (`chubchiks') in L\'evy walks [38]. They too vary similarly with the parameter $\mu$, which determines the run time PDF, $\psi(t) \propto t^{-1-\mu}$. For L\'evy walks in the superdiffusive case ($1<\mu<2$), the region near the initial position is Gaussian with the tails of the distribution $|x|> \nu t$ having the distribution $p(x,t) \sim t/|x|^{1+\mu}$. Although our correlated random walk has similarities to L\'evy walks, the major difference in the asymptotic density is the continuous distribution of the bimodal peaks at positions $|x|<\nu t$ instead of the chubchiks seen in L\'evy walks at $|x|=\nu t$.

\begin{figure}[h!]
	\includegraphics[width=\linewidth]{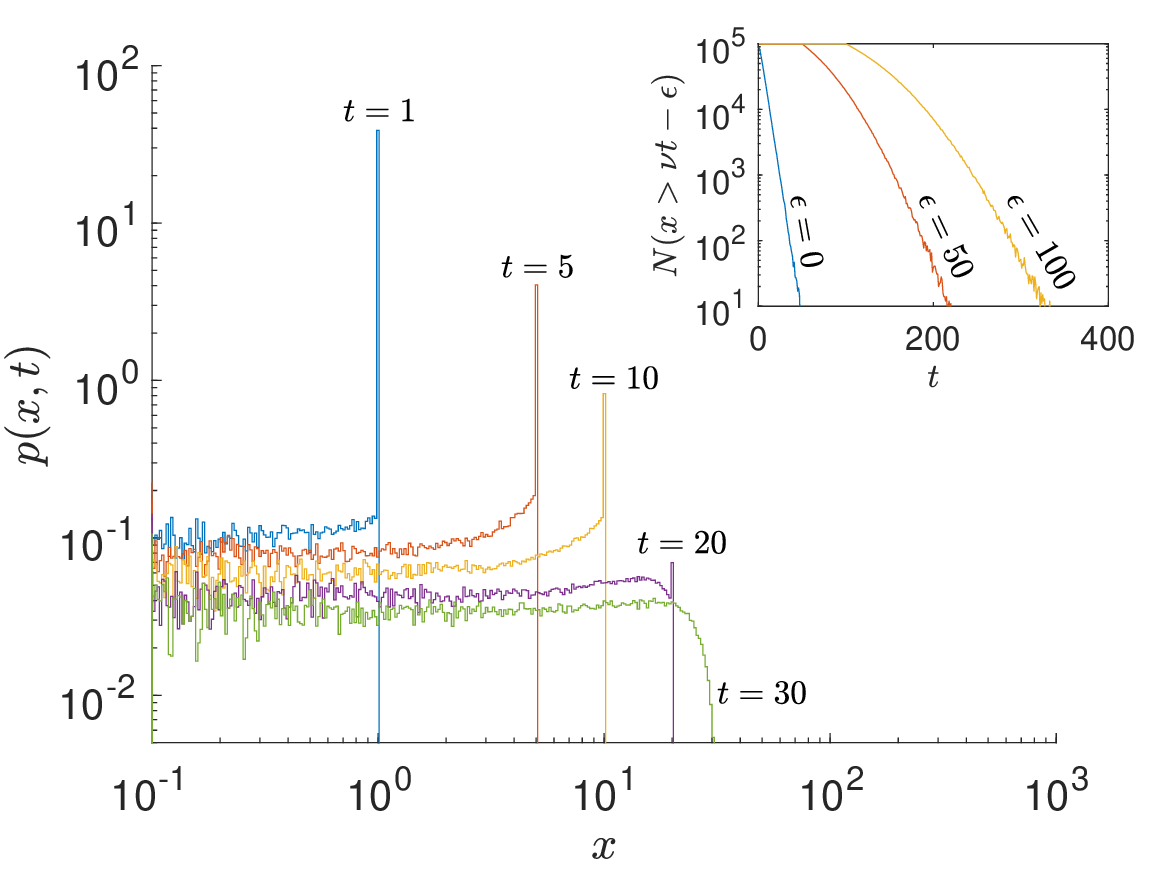}
	\caption{\textit{Main:} The PDF of particle positions, $P(x,t)$, for simulations of (9) at varying times with $w=0.8$. Other parameters are $N=10^5$, $u=0.5$, $\nu=1$ and $\lambda=1$ \textit{Inset:} From the same simulation as the main figure we count the number of particles, $N(x>\nu t-\epsilon)$, out of $N=10^5$ that have position $x>\nu t -\epsilon$ with $\epsilon$ varied between $0$, $50$ and $100$. The maximum position possible is $x=\nu t$. }
	\label{poshist_vartime}
\end{figure}

In essence, the chubchiks of L\'evy walks appear due to the group of particles that have been moving at the propagation velocity for the entire time $t$ and thus form a propagating front. Intriguingly, these fronts also appear for our correlated random walk but at very short times shown by Fig. \ref{poshist_vartime}. However, these propagating fronts decay exponentially with time whereas for L\'evy walks they decay as $t^{1-\mu}$ ($1<\mu<2$). By $t=30$, the propagating front has completely `evaporated' and the tail is now exponential with no trace of the original front. This phenomenon is intuitive since particles performing our correlated random walk take exponentially distributed runs, abeit in a persistent manner, but L\'evy walks take power-law distributed runs. Exponential decay of the fronts can be seen in the inset of Fig. \ref{poshist_vartime} where the number of particles $N(\cdot)$ with position $x>\nu t -\epsilon$ is plotted as a function of time $t$.

The evaporation of the propagating front demonstrates a non-equilibrium phase transition since the PDF shows chubchiks for short times that decay into exponential tails for long times. This shows the non-stationary nature of the random walk generated by (9) and the transition from L\'evy walk like behavior at short times to a completely novel distribution for long times.

\end{document}